\documentclass[12pt,a4paper]{article}
\usepackage[utf8]{inputenc}
\usepackage{epsfig}
\usepackage{mathtools}
\usepackage{amsfonts}
\usepackage{amssymb}
\usepackage{amsmath}
\usepackage{cancel}
\usepackage{color}
\usepackage{slashed}
\usepackage{cite}
\usepackage{caption}
\usepackage{subcaption}
\usepackage{comment}
\usepackage{marginnote}
\usepackage{mcite}

\allowdisplaybreaks

\def\bb {\begin {eqnarray}}
\def\ee {\end {eqnarray}}

\newcommand{\HS}{S}

\newcommand{\bea}{\begin{eqnarray}}
\newcommand{\eea}{\end{eqnarray}}
\newcommand{\z}{&&\hspace*{-1cm}}

\def\N {\mathcal{N}}

\def\({\left(}
\def\){\right)}

\def\P {{\bf P}}

\newcommand{\beq}{\begin{equation}}
\newcommand{\eeq}{\end{equation}}
\newcommand{\beqq}{\begin{equation*}}
\newcommand{\eeqq}{\end{equation*}}
\newcommand\beqa{\begin{eqnarray}}
\newcommand\eeqa{\end{eqnarray}}
\newcommand\beqaa{\begin{eqnarray*}}
\newcommand\eeqaa{\end{eqnarray*}}
\newcommand\beaa{\begin{array}}
\newcommand\eeaa{\end{array}}

\begin{document}
	
\begin{center}
		
\vspace{1cm}
		
{ \Large\bf DGLAP and BFKL equations in $\N=4$ SYM: \\ from weak to strong coupling} \vspace{1cm}

{\large A.V. Kotikov$^{1}$ and  A.I. Onishchenko$^{1,2}$}\vspace{0.5cm}
		
{\it $^1$Bogoliubov Laboratory of Theoretical Physics, Joint
Institute for Nuclear Research, Dubna, Russia, \\
$^2$Skobeltsyn Institute of Nuclear Physics, Moscow State University, Moscow, Russia}\vspace{1cm}

\abstract{DGLAP and BFKL equations are among the cornestones of the contemporary QCD. Moreover, they also played an important role in the recent studies of integrability structure of $\N = 4$ SYM. Here, we review the results obtained along this way together with a brief account of approaches and methods used.}

\end{center}

\tableofcontents{}\vspace{0.5cm}

\renewcommand{\theequation}{\thesection.\arabic{equation}}

\section{Introduction}

This paper deals with the review of the known properties of the 
Balitsky-Fadin-Kuraev-Lipatov (BFKL) 
 \cite{BFKL,*BFKL-1,*BFKL-2,*BFKL-3,*BFKL-4,*BFKL-5}
and Dokshitzer-Gribov-Lipatov-Altarelli-Parisi (DGLAP) \cite{DGLAP,*DGLAP-1,*DGLAP-2,*DGLAP-3,*DGLAP-4}
equations in   ${\mathcal N}=4$
Supersymmetric Yang-Mills (SYM) theory \cite{Brink:1976bc,Gliozzi:1976qd}.

Lev Lipatov is one of the main contributors to the discovery and subsequent study of both these well known equations. The BFKL and DGLAP equations resum, respectively, the most important  contributions proportional to
$\sim \alpha_s \ln(1/x)$ and $\sim \alpha_s \ln(Q^2/\Lambda^2)$ in two different
kinematical regions of the Bjorken parameter $x$ and the ``mass'' $Q^2$ of the
virtual photon for the process of deep inelastic lepton-hadron scattering (DIS) and as such, they are among the main ingredients in the analysis and description of experimental data for lepton-nucleon and nucleon-nucleon scattering.

In the case of supersymmeric theories these equations simplify drastically.
Moreover, in the case of $\N =4$ SYM they become related with each other for the nonphysical values of Mellin moments $j$ as it was proposed 
by Lev Lipatov in \cite{KL}. 

The first hint towards the simplification of BFKL and DGLAP equations in supersymmetric theories came from the study of  quasi-partonic operators in $\N = 1$ SYM in leading order (LO) approximation \cite{Bukhvostov:1985vj,Bukhvostov:1985rn}. It was shown, that quasi-partonic operators in this theory  are unified in supermultiplets and their anomalous dimensions could be described in terms of single universal anomalous dimension by shifting the argument of the latter by some integer number.
Calculations in ${\mathcal N}=4$ SYM, where the
coupling constant is not renormalized, gave even more remarkable results. Namely, it turned out, that all twist-2 operators in this theory belong to
the same supermultiplet with their anomalous dimension matrix fixed completely
by the superconformal invariance.   The  LO universal anomalous dimension in $\N = 4$ SYM was found to be proportional to $\Psi (j-1)-\Psi (1)$,
which implies that DGLAP evolution equations for the matrix elements of quasi-partonic operators  are equivalent to the Schr\"{o}dinger equation for
the integrable Heisenberg spin model~\cite{N=4,LN4}. In QCD the
integrability remains only in a small sector of these operators~\cite{Braun:1998id,Belitsky:1999qh} (see also~\cite{Ferretti:2004ba,Beisert:2004fv}). On the other hand, in the case of $\N =4$ SYM the equations for other sets of operators are also
integrable~\cite{Minahan:2002ve,Beisert:2003yb,Beisert:2003tq}.

Similar results related to integrability of BFKL and Bartels-Kwiecinski-Praszalowicz (BKP) \cite{BKP,*BKP-1} equations in the multi-colour limit of QCD were obtained earlier in~\cite{Integr}. Later it was also shown~\cite{KL}, that in the case of ${\mathcal N}=4$ SYM there is a deep relation between the BFKL and DGLAP evolution equations. Namely, the $j$-plane singularities of the anomalous dimensions of Wilson
twist-2 operators in this case can be obtained from the eigenvalues of the
BFKL kernel by analytic continuation. The next-to-leader (NLO)
calculations in $\N=4$ SYM demonstrated~\cite{KL}, that some of these
relations are valid also in higher orders of perturbation theory. In
particular, the BFKL equation has the property of the hermitian
separability and the eigenvalues of the
anomalous dimension matrix may be expressed in terms of the universal
function $\gamma _{uni}(j)$.

In what follows, in Section 2 we discuss the BFKL equation and its Pomeron solution both at weak and strong coupling values. Section 3 is devoted
to DGLAP equation and related anomalous dimensions of Wilson twist-2 operators. 
Here, as in a case of BFKL equation we are not limiting ourselves by weak coupling regime and discuss also the behavior of anomalous dimensions at strong coupling. The strong coupling study both in the case of BFKL and DGLAP equations was made possible by an extensive use of both integrability and AdS/CFT correspondence approaches. Finally, we come with a summary in  Conclusion.

\section{BFKL equation and Pomeron}

The behavior of scattering amplitudes in high energy or Regge limit may be conveniently described by the positions of singularities of their partial wave amplitudes in complex angular momentum plane. In particular, the behavior of the total cross-section for the scattering of colorless particles at high energies is related to the Regge pole with  quantum numbers of the vacuum and even parity - the so called Pomeron. Within perturbation theory the latter is given by the bound state of two reggeized gluons and is described by now famous BFKL 
%Balitsky-Fadin-Kuraev-Lipatov (BFKL) 
equation  \cite{BFKL}. In what follows we are going to describe the properties of BFKL equation and its Pomeron solution in the case of maximally supersymmetric $\N = 4$ Yang-Mills theory.

\subsection{BFKL equation and perturbation theory}

In the high-energy limit ($s\gg -t$) the total cross-section $\sigma (s)$ for the  scattering of
colourless particles $A$ and $B$ takes the following factorized form
\begin{equation}
\sigma (s)~=~\int \frac{d^{2}q\,d^{2}q^{\prime }}{(2\pi
	)^{2}\,q^{2}\,q^{\prime 2}}\Phi _{A}(q)\,\Phi _{B}(q^{\prime
})\int_{a-i\infty }^{a+i\infty }\frac{d\omega }{2\pi i}\left(
{\frac{s}{s_0}}\right) ^{\omega }G_{\omega }(q,q^{\prime }), ~~~
s_0 = |q||q^{\prime }|\, ,
\label{BFKL}
\end{equation}
where  $\Phi_{i}(q_{i})$ are impact factors of the colliding particles with momenta $p_A$, $p_B$, $s=2p_{A}p_{B}$ is their squared invariant mass and $G_{\omega }(q,q^{\prime })$ is the $t$-channel partial wave for the reggeized gluon-gluon scattering. The latter depends on two transverse momenta of the reggeized gluons in the $t$ channel  $q$ and $q^{\prime }$\footnote{Here and below we omit arrows in the notation of transverse momenta and simply write $q$ and $q'$ instead of $\vec{q}$ and $\vec{q^{\prime}}$.} and is given by the solution to  BFKL equation. 

Using the dimensional regularization and the $\overline{MS}$-scheme to deal with ultraviolet and infrared divergences in the
intermediate expressions, the BFKL equation for $G_{\omega }(q,q^{\prime })$
can be written in the form
\begin{equation}
\omega G_{\omega }(q,q_1)~=~\delta ^{D-2}(q-q_{1})+\int
d^{D-2}q_{2}\,K(q,q_{2})\,G_{\omega }(q_{2},q_{1})\,,
\end{equation}
where
\begin{equation}
K(q_{1},q_{2})~=~2\,\omega (q_{1})\,\delta
^{D-2}(q_{1}-q_{2})+K_{r}(q_{1},q_{2})
\end{equation}
and the space-time dimension $D=4-2\varepsilon $ for $\varepsilon \to 0$. Here,  
$\omega (q)$ is the gluon Regge trajectory and the integral kernel $%
K_{r}(q_{1},q_{2})$ is related to the real particle production in $t$-channel. In $\N = 4$ SYM the description of BFKL equation follows closely the one in the case of QCD. Moreover, in leading logarithmic approximation the expressions for BFKL integral kernel in QCD and $\N =4$ SYM coincide and the difference starts at the level of NLO corrections. Initially NLO radiative  corrections to  BFKL integral kernel at $t=0$ were calculated in the case of QCD in \cite{next,*next-1}
\footnote{The $t\neq 0$ case can be found in the recent papers \cite{Fadin:2007xy}.} and later were generalized to the case of $\N = 4$ SYM in \cite{KL00}.

As it was shown in \cite{next,KL00}, a complete and orthogonal set of
eigenfunctions of the homogeneous BFKL equation in LO
%the leading order (LO) 
is given by ($\gamma_{BFKL} = \frac{1}{2} + i\nu$):
\begin{equation}
G_{n,\gamma }(q^2/q^{\prime 2},\theta )~=~\left( \frac{q^{2}}{q^{\prime 2}}
\right) ^{\gamma_{BFKL} -1}e^{in\theta }\, ,
\end{equation}
where integer conformal spin $n$ and real $\nu$ enter the parametrization of conformal weights $m$ and $\tilde m$ of the principal series of unitary M\"{o}bius group representation
\begin{equation}
m = \frac{1+n}{2} + i\nu\, , \quad \tilde{m} = \frac{1-n}{2} + i\nu \label{conformal-weights}
\end{equation}

The BFKL kernel in this representation is diagonalized up to the effects
related with the running coupling constant $a_{s}(q^{2})$ and up to NLO in the case of QCD we have
\bea
\omega^{QCD}_{\overline{MS}} = 4 a_{s}(q^{2})\biggl[
\chi (n,\gamma_{BFKL})+\delta^{QCD}_{\overline{MS}} (n,\gamma_{BFKL} )
a_{s}(q^{2}) \biggr] \,, 
\label{ome1a}
\eea
where LO and NLO eigenvalues of BFKL kernel using formulae of \cite{KL00} are given by
\begin{eqnarray}
\chi (n,\gamma ) &=&2\Psi (1)-\Psi \Bigl(\gamma +\frac{n}{2}\Bigr)-\Psi
\Bigl%
(1-\gamma +\frac{n}{2}\Bigr)  \label{7} \\
&&  \nonumber \\
%\begin{eqnarray}
\delta^{QCD}_{\overline{MS}} (n,\gamma ) &=&
\biggl(\frac{67}{9} -2\zeta (2)
-\frac{10}{9} \,
%\hat{n}_f
\frac{n_{f}}{N_{c}}
%-\frac{4}{9} \, \hat{n}_s
%\frac{n_{s}}{N_{c}}
\biggr) %
\chi (n,\gamma )
%+6\zeta (3)
+6\zeta (3)+
\Psi ^{\prime \prime }\Bigl(\gamma +\frac{n}{2}\Bigr)+ \Psi ^{\prime
	\prime } \Bigl(1-\gamma +\frac{n}{2}\biggr)
\nonumber \\
&-&  2\Phi (n,\gamma )-2\Phi (n,1-\gamma )
-\biggl(\frac{11}{3}-
\frac{2}{3} \,
%\hat{n}_f
\frac{n_{f}}{N_{c}}
\biggr) \frac{1}{2}
%\biggl(
\chi ^{2}(n,\gamma )
\nonumber \\
&+&\frac{\pi ^{2}\cos (\pi \gamma )}{\sin ^{2}(\pi \gamma
	)(1-2\gamma )}\Biggl\{
\biggl(1
%-\overline{n}_f + \frac{\overline{n}_s}{2}
+\frac{\tilde{n}_{f}}{N_{c}^{3}}
%-\frac{\tilde{n}_{s}}{2 N_{c}^{3}}
\biggr) \frac{\gamma (1-\gamma )}{
	2(3-2\gamma )(1+2\gamma )}\cdot \delta _{n}^{2}
\nonumber \\
&-& \biggl(3+\biggl(1
%- \overline{n}_f
+\frac{\tilde{n}_{f}}{N_{c}^{3}}
\biggr)
\frac{2+3\gamma (1-\gamma )}{(3-2\gamma )(1+2\gamma )}
\biggr)\cdot \delta_{n}^{0}
\Biggr\},
\label{8.2}
\end{eqnarray}
Here, 
$\delta _{n}^{m}$ is the Kroneker symbol, and $\Psi (z)$, $\Psi
^{\prime }(z)$ and $\Psi ^{\prime \prime }(z)$ are the Euler $\Psi $
-function and its  derivatives. The function $\Phi (n,\gamma
)$ is given by
\begin{eqnarray}
\z \Phi (n,\gamma ) ~=~
~\sum_{k=0}^{\infty }\frac{(-1)^{k+1}}{%
	k+\gamma +n/2}\Biggl[ \Psi ^{\prime }(k+n+1)-\Psi ^{\prime }(k+1)
\nonumber \\
\z
+(-1)^{k+1}
\Bigl(\beta ^{\prime }(k+n+1)+\beta ^{\prime }(k+1)\Bigr)
%\nonumber \\ &-&
- \frac{1}{k+\gamma +n/2}\biggl( \Psi (k+n+1)-\Psi (k+1)%
\biggr) \Biggr] \nonumber \\  \label{9}
\end{eqnarray}
and
\[
\beta ^{\prime }(z)=\frac{1}{4}\Biggl[ \Psi ^{\prime }\Bigl(\frac{z+1}{2}%
\Bigr)-\Psi ^{\prime }\Bigl(\frac{z}{2}\Bigr)\Biggr]
\]
To obtain the corresponding results in $\N = 4$ SYM we should account for contributions of scalars and fermions transforming in the adjoint representation of the gauge group. This way in $\overline{DR}$ scheme \cite{DRED} we get
%(see \cite{KoLi00,KoLi03}):
\begin{eqnarray}
\delta ^{N=4}_{\overline{DR}}(n,\gamma )&=&
6\zeta (3)+
\Psi ^{\prime \prime }\Bigl(\gamma +\frac{n}{2}\Bigr)+ \Psi ^{\prime
	\prime } \Bigl(1-\gamma +\frac{n}{2}\biggr)
\nonumber \\
&-&  2\Phi (n,\gamma )-2\Phi (n,1-\gamma ) -
2\zeta (2) \chi (n,\gamma ),   \label{K3}
\end{eqnarray}
where the $\overline{DR}$
%corresponding 
coupling constant $\hat{a}_{s}$ is related  to $\overline{MS}$ one  $a_{s}$ as \cite{Altarelli,*Altarelli-1}
\begin{eqnarray}
\hat{a}_{s} ~=~ a_{s} + \frac{1}{3} a^2_{s}.
\label{9aa}
\end{eqnarray}
It should be noted, that the sum $\Phi (n,\gamma )+\Phi (n,1-\gamma )$
can be further rewritten (see \cite{KL,*KL-1}) as a combination of functions with
arguments dependent on $\gamma +n/2 \equiv M$ and
$1-\gamma +n/2 \equiv \tilde{M}$ only. Indeed, we have
\begin{eqnarray}
&&\Phi (n,\gamma )+\Phi (n,1-\gamma )
%\Phi (|n|,\gamma )+\Phi (|n|,1-\gamma )
=\chi (n,\gamma )\,\left( \beta
^{\prime }(M)+\beta ^{\prime }(1-\widetilde{M})\right)   \nonumber \\
&&+\Phi _{2}(M)-\beta ^{\prime }(M)\left[ \Psi (1)-\Psi (M)\right] +\Phi
_{2}(1-\widetilde{M})-\beta ^{\prime }(1-\widetilde{M})\left[ \Psi (1)-\Psi (1-%
\widetilde{M})\right] ,  \nonumber
\end{eqnarray}
where $\chi (n,\gamma )$ is given by Eq. \eqref{7} and
\begin{eqnarray}
\Phi_2 (M )=~\sum_{k=0}^\infty
\frac{\left( \beta ^{\prime }(k+1)+
	(-1)^k \Psi ^{\prime }(k+1)\right) }{k+M}
-\sum_{k=0}^\infty \frac{(-1)^k\left( \Psi (k+1)-\Psi (1)\right) }{%
	(k+M)^2} \,.
\label{9.1}
\end{eqnarray}
From the latter property we see, that BFKL equation in $\N = 4$ SYM satisfies the property of  hermitian separability (see Ref. \cite{KL} and discussions therein). Moreover, the NLO eigenvalue of BFKL kernel in $\N = 4$ SYM could be further rewritten in terms of analytically continued harmonic sums, see for example appendix C in \cite{BFKLQSC3}. This way we first get the property of maximal transcedentality\footnote{See the discussion for the case of anomalous dimensions in the next section} \cite{KL} manifest directly\footnote{Actually it could be already  inferred  from the expression in Eq. \eqref{K3}, see discussion in \cite{KL}} for the eigenvalues of BFKL kernel expressed in terms of harmonic sums with well defined weights and second - the expressions for Pomeron intercept in $\N = 4$ SYM as a function of conformal spin $n$ take the form \cite{BFKLQSC3}:
\begin{align}
\omega_0^{LO} &=4\mathbb{S}_1\left(\frac{n-1}{2}\right)\, , \\
\omega_0^{NLO} &= 8\mathbb{S}_{2,1}\left(\frac{n-1}{2}\right)+8\mathbb{S}_3\left(\frac{n-1}{2}\right) +\frac{4\pi^2}{3}\mathbb{S}_1\left(\frac{n-1}{2}\right)\, , 
\end{align}
where $\omega_0 = \omega (n,1/2)$ and
\beq\label{binomial}
\mathbb{S}_{i_1,\dots,i_k}(M)=(-1)^M \sum\limits_{j=1}^M(-1)^j \(\begin{array}{c}
M\\
j
\end{array}\)
\(\begin{array}{c}
M+j\\
j
\end{array}\)
S_{i_1,\dots,i_k}(j)\;.
\eeq 
are binomial harmonic sums \cite{Vermaseren:1998uu}.

\subsection{AdS/CFT correspondence and Pomeron at strong coupling}

AdS/CFT-correspondence
\cite{AdS-CFT,AdS-CFT1,Witten} provides us with a unique possibility to study BFKL equation and its Pomeron solution at strong coupling.  In the framework of AdS/CFT correspondence the BFKL Pomeron is equivalent to the reggeized graviton
\cite{Polchinski:2002jw}, what in particular allows us to construct the Pomeron interaction model as a generally covariant effective theory for the reggeized gravitons \cite{Lipatov:2011ab}. The Pomeron intercept for zero conformal spin $n=0$ and at the leading order in the inverse coupling constant could be easily obtained using diffusive approximation for BFKL equation \cite{KL00,Fadin:2007xy}, strong coupling results
for anomalous dimensions \cite{Gubser:2002tv,Frolov:2002av,Roiban:2007jf} 
and the mentioned Pomeron-graviton duality 
\cite{Polchinski:2001tt,Polchinski:2002jw}. This is the approach, which was used in\footnote{See Erratum to that paper} \cite{KLOV}, see also \cite{Brower:2006ea} for similar results. Moreover, it turns out that the approach of \cite{KLOV} could be also extended to include higher order corrections in inverse coupling constant \cite{Kotikov:2013xu}.  First, one should note that due to symmetry of BFKL kernel eigenvalue in $\N = 4$ SYM under substitution $\gamma_{BFKL}\to 1 - \gamma_{BFKL}$ the latter is an even function of $\nu$:
\beq
j-1 = \omega = \omega_0 + \sum_{m=1}^{\infty} (-1)^m \, D_m \, \nu^{2m} \, ,
\label{6i}
\eeq
where ($\lambda = g^2 N_c$ is t'Hooft coupling constant)
\bea
\omega_0 &=& 4\ln 2 \, \frac{\lambda}{4\pi^2} \left[ 1- \overline{c}_1
\frac{\lambda}{16\pi^2}  \right] +
O(\lambda^3) \,, \label{7i} \\
D_m &=&
%(-1)^m \,
2\left(2^{2m+1}-1\right)\zeta_{2m+1} \frac{\lambda}{4\pi^2} +\frac{\delta ^{(2m)}(1/2)}{(2m)!}\,
\frac{\lambda ^2}{64 \pi ^4}+
O(\lambda^3) \,.
\label{8i}
\eea
and according to \cite{KL} we have
\beq
%\hat{\delta}_1
\overline{c}_1 ~=~ 2 \zeta_2 +
\frac{1}{2\ln 2} \left(11\zeta_3
- 32{\rm Ls}_{3}\Bigl(\frac{\pi }{2}\Bigl) -14 \pi \zeta_2
\right) \approx  7.5812 \,, ~~~
%\label{8.1i}
%\ee
%where (see \cite{Lewin})
%%\cite{Lewin,Devoto})
%\be
{\rm Ls}_{3}(x)=-\int_{0}^{x}\ln ^{2}\left| 2\sin \Bigl(\frac{y}{2}%
\Bigr)\right| dy \,. \label{8i}
\eeq
Second, the M\"{o}bius invariance and hermicity of BFKL kernel suggest that the expansion \eqref{6i} should be also valid at strong coupling. Next, to make the connection with the results for anomalous dimensions of local composite operators\footnote{These are anomalous dimensions of twist 2 operators contributing to DGLAP equation \cite{DGLAP,*DGLAP-1,*DGLAP-2,*DGLAP-3,*DGLAP-4}} (string energies in the language of AdS/CFT correspondence) $\gamma$ at strong coupling we use the relation of the latter to BFKL anomalous dimension $\gamma_{BFKL}$ \cite{next,Salam:1998tj,Salam:1999cn}:  
\beq
\gamma ~=~ \gamma_{BFKL} + \frac{\omega}{2} ~=~ \frac{j}{2}+i\nu \,.
%\gamma_{BFKL} + \frac{\omega}{2} ~=~ \gamma_{BFKL} + \frac{j-1}{2},
\label{2e}
\eeq
Now, due to the energy-momentum conservation $\gamma (j=2) = 0$ we have 
\beq
\nu(j=2)=i\,
\label{4e}
\eeq
and the small $\nu$ expansion for the eigenvalues of the BFKL kernel for reggeized graviton takes the form 
\beq
j-2 = \sum_{m=1}^{\infty}  D_m \left({(-\nu^{2})}^m-1\right),
\label{7e}
\eeq
where $\nu^2$ is related to $\gamma$ according to Eq. \eqref{2e} 
\beq
\nu^2=
-{\left(\frac{j}{2}
	-\gamma\right)}^2.
\label{8e}
\eeq
On the other hand, due to the
ADS/CFT correspondence the string energies $E$ in dimensionless units are related to the anomalous dimensions $\gamma$ of  twist-two
operators as\footnote{Note that our expression (\ref{1a}) for the  string energy $E$ differs from a definition, in which $E$ is equal to the scaling dimension $\Delta_{sc}$. Still the equation \eqref{1a} is correct as it can be written as $E^2=(\Delta_{sc}-2)^2-4$ and coincides with Eqs. (45) and (3.44) from Refs. \cite{AdS-CFT1} and \cite{Witten}, respectively.} 
\cite{AdS-CFT1,Witten}
\begin{equation}
E^2=(j+\Gamma)^2-4,~~\Gamma=-2\gamma
\label{1a}
\end{equation}
and therefore we can obtain from Eq. (\ref{8e}) the relation between the parameter $\nu$ of the conformal  weights for
the principal series of unitary representations of the M\"{o}bius group and the string energy $E$
\beq
\nu^2=
-\left(\frac{E^2}{4} +1\right)\,.
\label{3a}
\eeq
This expression for $\nu ^2$ can
%should
be inserted in the r.h.s. of Eq. (\ref{7e}) and gives us the following expression for the Regge trajectory
of the graviton in the AdS space
\beq
j-2 =  \sum_{m=1}^{\infty}
D_m \left[{\left(\frac{E^2}{4}+1\right)}^m-1\right].
\label{4a}
\eeq
Next, we assume that Eq. \eqref{4a} is also valid both  at large $j$ and large $\lambda$ in the region $1\ll j \ll\sqrt{\lambda}$,
where the calculations of string energies at strong coupling were performed~\cite{Gromov:2011de,Roiban:2011fe}.
These energies can be presented  in the form
\footnote{Here we put $S=j-2$, which in particular is related to the
	use of the angular momentum $J_{an}=2$ in calculations of Refs \cite{Gromov:2011de,Roiban:2011fe}.}
\beq
\frac{E^2}{4} ~=~ \sqrt{\lambda} \, \frac{S}{2}\, \left[h_0(\lambda)
+ h_1(\lambda) \frac{S}{\sqrt{\lambda}} + h_2(\lambda) \frac{S^2}{\lambda}
\right] + O\Bigl(S^{7/2}\Bigr),
\label{5a}
\eeq
where
\beq
h_i(\lambda) ~=~  a_{i0} + \frac{a_{i1}}{\sqrt{\lambda}} +
\frac{a_{i2}}{\lambda} +  \frac{a_{i3}}{\sqrt{\lambda^3}} +
\frac{a_{i2}}{\lambda^2}.
\label{5.1a}
\eeq
The contribution proportional to $S$ can be extracted from the result of Basso
\cite{Basso:2011rs,Basso:2012ex} taking $J_{an}=2$ according to \cite{Gromov:2011bz}:
\beq
h_0(\lambda) =
\frac{I_3(\sqrt{\lambda})}{I_2(\sqrt{\lambda})} + \frac{2}{\sqrt{\lambda}} =
\frac{I_1(\sqrt{\lambda})}{I_2(\sqrt{\lambda})} - \frac{2}{\sqrt{\lambda}}\, ,
\label{Ad5.1}
\eeq
where $I_k(\sqrt{\lambda})$ is the modified Bessel functions. This gives us the 
following values of coefficients $a_{0i}$
\beq
a_{00} ~=~ 1,~~ a_{01}~=~ - \frac{1}{2},~~a_{02} ~=~ a_{03}~=~  \frac{15}{8},~~
a_{04}~=~  \frac{135}{128}
\label{Ad5.2}
\eeq
Next, the coefficients $a_{10}$ and $a_{20}$ come from the consideration of the
classical energy of the folded spinning string configurations corresponding to the twist-two operators (see, for example, \cite{Roiban:2011fe})
\beq
a_{10}~=~  \frac{3}{4},~~a_{20} ~=~ - \frac{3}{16}\,.
\label{Ad5.3}
\eeq
Finally, the one-loop coefficient $a_{11}$ was found  in \cite{Gromov:2011bz}
considering different asymptotical regimes  and accounting for Basso result  \cite{Basso:2011rs}
\beq
a_{11}~=~ \frac{3}{16} \Bigl(1-\zeta_3\Bigr),
\label{Ad5.4}
\eeq
Now, comparing both sides of Eq. \eqref{4a} at large $j$ values gives us the desired values of  coefficients $D_m$ (see Appendix A in \cite{Kotikov:2013xu}) and the value for Pomeron intercept $j_0$ at zero conformal spin in the limit $\lambda\to\infty$ takes the form
\footnote{Using a similar approach, the coefficients $\sim \lambda^{-1}$ and 
$\sim \lambda^{-3/2}$ were calculated also in the paper \cite{Costa:2012cb}.}
%(see Appendix (\ref{11.dc3}))
\beq
j_0 ~=~ 2 -  \frac{2}{\lambda^{1/2}} \, \left[1
+
%\Bigl(8\ln 2 -3\Bigr) \,
\frac{1}{2\lambda^{1/2}} - \frac{1}{8\lambda} -
\Bigl(1 +3 \zeta_3 \Bigr) \frac{1}{\lambda^{3/2}} +
%\left(1-%\frac{33}{4}-36\ln 2 + 48 \ln^2 2 - 2 a_{02} \right)
\left(2a_{12} - \frac{145}{128} - \frac{9}{2}\zeta_3 \right)
\frac{1}{\lambda^2} + O\left(\frac{1}{\lambda^{5/2}}\right)
\right] \, . \label{11.de}
\eeq
with unknown coefficient $a_{12}$ in $1/\lambda^2$ correction in Eq. \eqref{11.de}. Later the strong coupling expansion of Pomeron intercept for zero conformal spin was calculated even further within quantum spectral curve (QSC) approach with the result given by \cite{QSCwork}:
\beqa 
j_0 =  &=& 2 -\frac{2}{\lambda^{1/2}}-\frac{1}{\lambda }+ \frac{1}{4\,\lambda^{3/2}}+\left(6 \zeta_3+2\right) \frac{1}{\lambda^2} \\
&+& \left(18 \, \zeta_3 + \frac{361}{64} \right) \frac{1}{\lambda^{5/2}} + \left(39 \, \zeta_3 + \frac{511}{32}\right) \frac{1}{\lambda^3}  + \mathcal{O}\left(\frac{1}{\lambda^{7/2}}\right).
\eeqa

\subsection{BFKL equation and integrability}

The most intriguing property of BFKL equation in $\N = 4$ SYM is its all-loop integrability tightly connected to all-loop integrability of anomalous dimensions of local operators in $\N = 4$ SYM, see \cite{integrability-review} for a review. Initially, the integrability of multicolor QCD was discovered in LO by Lev Lipatov precisely in the study of Regge limit of high-energy scattering \cite{Integr,*Integr-1} and only latter in the study of anomalous dimensions of composite operators \cite{N=4,LN4}. Calculations in the maximally supersymmetric $\N=4$ SYM, where the coupling constant is not renormalized and all twist-2 operators enter in the same multiplet, showed  that their anomalous dimension matrix is fixed completely by the super-conformal invariance and its entries are expressed in terms of universal anomalous dimension, which in LO is proportional to $\Psi (S-1)-\Psi (1)$ (see the discussion in next section). The latter property means, that the evolution equations for the matrix elements of quasi-partonic operators in the multicolour limit $N_{c}\rightarrow \infty $ are equivalent to
the Schr\"{o}dinger equation for an integrable Heisenberg spin model \cite{N=4,LN4}. Subsequent study of these and other operators  both at leading   and higher orders, starting from the works of \cite{MinahanZarembo,N4SuperSpinChain} showed that the discovered integrability property in $\N = 4$ SYM may be extended to all loops. It turned out, that a lot of methods developed for two-dimensional integrable systems, such as sigma-model and spin-chain S-matrices \cite{StaudacherSMatrix,BetheAnsatzQuantumStrings,BeisertDynamicSmatrix,BeisertAnalyticBetheAnsatz,TranscedentalityCrossing,JanikWorldsheetSmatrix,ArutyunovFrolovSmatrix,ZamolodchikovFaddevAlgebraAdS}, Asymptotic Bethe Ansatz (ABA) \cite{MinahanZarembo,N4SuperSpinChain,LongRangeBetheAnsatz,TranscedentalityCrossing}, Thermodynamic Bethe Ansatz (TBA) \cite{TBAN4,TBAN4proposal,TBAexcitedstates,TBAMirrorModel} as well as $Y$ and $T$-systems \cite{YsystemAdS5,TBAfromYsystem,WronskianSolution,SolvingYsystem} could be also used for computation of anomalous dimensions of $\N = 4$ SYM operators not only at weak but also strong and in general finite coupling constant.  The most advanced framework for such spectral problem calculations at the moment is offered by quantum spectral curve (QSC) method. The latter is an alternative reformulation of TBA equations as a finite set of algebraic relations for Baxter type $Q$-functions together with analyticity and Riemann-Hilbert monodromy conditions for the latter \cite{N4SYMQSC1,N4SYMQSC2,twistedN4SYMQSC,N4SYMQSC3,N4SYMQSC4}. Within the quantum spectral curve formulation one can relatively easy obtain numerical solution for any coupling and state \cite{QSCnumericsN4SYM1,QSCnumericsN4SYM2}. Also, QSC formulation allowed to construct iterative perturbative solutions for these theories at weak coupling up to, in principle, arbitrary loop order \cite{VolinPerturbativeSolution,N4SYMQSC4}. What is more important QSC allows also for the study of BFKL regime \cite{BFKLQSC1,BFKLQSC2,BFKLQSC3}.

The integrability of LO BFKL equation (in leading logarithmic approximation) could be conveniently studied in the impact parameter space $\vec{\rho}$, where the BFKL equation has the Schro\"odinger like form \cite{Integr}:
\begin{equation}
E f(\vec{\rho_1},\vec{\rho_2}) = H_{12} f(\vec{\rho_1}, \vec{\rho_2})\, 
\end{equation}
where in leading logarithmic approximation hamiltonian $H_{12}$ takes holomorphically separable form
\begin{eqnarray}
H_{12} &=& h_{12} + h^*_{12}\, \nonumber \\
h_{12} &=& \ln (p_1 p_2) + \frac{1}{p_1} (\ln\rho_{12})p_1 + \frac{1}{p_2} (\ln\rho_{12})p_2 + 2 \gamma_E\, .
\end{eqnarray} 
Here impact parameters are complex coordinates $\rho_k = x_k + i y_k$, $\rho_{12} = \rho_1 - \rho_2$ and $p_k$, $p^*_k$ play the role of their canonically conjugated momenta. Moreover, it turns out that hamiltonian $H_{12}$ coincides with the hamiltonian of Heisenberg (XXX) $SL(2,C)$ spin chain \cite{Integr}, where $SL(2,C)$ (M\"obius) group generators are given by 
\begin{equation}
M_k^3 = \rho_k\partial_k\, , \quad M_k^+ = \partial_k\, , \quad M_k^- = -\rho_k^2\partial_k
\end{equation}
and $\partial_k = \partial/(\partial\rho_k)$. 

The eigenfunctions of LO BFKL hamiltonian are easy to find with the use of M\"obius invariance. This way we get wave functions labeled by two conformal weights \eqref{conformal-weights} \cite{BFKLLOwavefunction}: 
\begin{eqnarray}
f_{m,\tilde{m}} (\vec{\rho}_1, \vec{\rho}_2 ; \vec{\rho}_0) &=& \left(
\frac{\rho_{12}}{\rho_{10}\rho_{20}}
\right)^m \left(
\frac{\rho^*_{12}}{\rho^*_{10}\rho^*_{20}}
\right)^{\tilde{m}}\, , 
\end{eqnarray}
with eigenvalues
\begin{equation}
E_{m,\tilde{m}} = 4 \text{Re} \left[\Psi \left(\frac{1+n}{2}+\i\nu\right)\right] - 4\Psi (1)\, .
\end{equation}
The Pomeron intercept as a function of conformal spin $n$ is then given by\footnote{$\lambda = g^2 N_c$ is again t'Hooft coupling constant.} 
\begin{equation}
\omega_0 = -\frac{\lambda}{8\pi^2} E_{m,\tilde{m}}\Big|_{\nu=0}
\end{equation}
Having established a connection of LO BFKL hamiltonian with integrable $SL(2,C)$ spin chain, we may write the generating function for its integrals of motion $q_r$ in terms of spin chain transfer matrix $T(u)$ \cite{Integr}:
\begin{equation}
T(u) = \text{Tr} \left(
L_1 (u) L_2 (u) \ldots L_n (u) 
\right) = \sum_{r=0}^n u^{n-r} q_r\, ,
\end{equation}
where Lax $L_k$ - operators are given by 
\begin{equation}
L_k (u) = \begin{pmatrix}
u+\rho_k p_k & p_k \\
-\rho_k^2 p_k & u - \rho_k p_k
\end{pmatrix} = 
\begin{pmatrix}
u & 0 \\ 0 & u
\end{pmatrix} + \begin{pmatrix}
1 \\ - \rho_k
\end{pmatrix} \begin{pmatrix}
\rho_k & 1
\end{pmatrix} p_k\, .
\end{equation}

To study BFKL equation within integrability approach beyond LO it is convenient to use quantum spectral curve method \cite{N4SYMQSC1,N4SYMQSC2,twistedN4SYMQSC,N4SYMQSC3,N4SYMQSC4}. Besides reproducing LO BFKL eigenvalues \cite{BFKLQSC1} authors of \cite{BFKLQSC2} were able to deduce the next-to-next-to-leading order (NNLO)
%NNLO 
corrections for BFKL eigenvalues at zero conformal spin in agreement with \cite{NNLO-BFKL-Velizhanin,Caron-Huot:2016tzz}. In addition, the authors of \cite{BFKLQSC3} obtained NNLO corrections for Pomeron intercept as a function of conformal spin $n$:
\beq\label{NNLO_intercept_function}
\omega_0^{NNLO} =32\(\mathbb{S}_{1,4}-\mathbb{S}_{3,2}-\mathbb{S}_{1,2,2}-\mathbb{S}_{2,2,1}-2 \mathbb{S}_{2,3}\)-\frac{16\pi^2}{3} \mathbb{S}_3-\frac{32\pi^4}{45} \mathbb{S}_1\; ,
\eeq
where the arguments of the binomial sums \eqref{binomial} are $(n-1)/2$. There are also partial results for the next-to-next-to-next-to-leading order (NNNLO)
%NNNLO 
intercept \cite{BFKLQSC3}. 

The range of applicability of QSC approach is not limited by weak coupling expansions but could be also used to obtain results at strong coupling. This way for Pomeron intercept  as a function of conformal spin $n$ we get \cite{BFKLQSC3}:
\begin{multline}
j_0=2-n
+\frac{(n-1)(n+2)}{{\lambda}^{1/2}}- \\
-\frac{(n-1)(n+2)(2n-1)}{2{\lambda}}
+\frac{(n-1)(n+2)(7n^2-9n-1)}{8{\lambda}^{3/2}}+\mathcal{O}\(\frac{1}{\lambda^2}\),
\end{multline}  
There are also analytical results at finite values of coupling constant for slope to intercept and curvature functions \cite{BFKLQSC3}. Finally, it should be mentioned that numerically QSC approach allows study of BFKL eigenvalues at arbitrary values of coupling constant without restrictions to weak or strong coupling regime considered above, see \cite{BFKLQSC3} for more details.

\section{Anomalous dimensions and DGLAP equation}

Phenomenologically, Bjorken scaling violation for parton distributions\footnote{Here we consider only the spin-average case. The
results for the spin-dependent case can be found in \cite{KL}.}	
in a framework of QCD  is governed by the anomalous dimensions matrix $\gamma _{ab}(j)$
%(the symbol $\tilde{} $ is used for spin-dependent case and 
(here $a_s=\alpha_s/(4\pi)$)
\begin{eqnarray}
\gamma _{ab}(j) &=& \int_{0}^{1}dx\,\,x^{j-1} W_{b\rightarrow a}(x)
~=~ \sum_{k=0}^{\infty} \gamma^{(k)} _{ab}(j) a_s^{k+1} 
%+ \gamma^{(1)} _{ab}(j) a_s^2+\gamma^{(2)} _{ab}(j) a_s^3 + O(a_s^4)
,~~
%\nonumber \\
%\tilde{\gamma} _{ab}(j) &=& \int_{0}^{1}dx\,\,x^{j-1}
%\tilde{W}_{b\rightarrow a}(x)
%~=~ \sum_{k=0}^{\infty} \tilde{\gamma}^{(k)} _{ab}(j) a_s^{k+1} 
%%+ \tilde{\gamma}^{(1)} _{ab}(j) a_s^2+
%%\tilde{\gamma}^{(2)} _{ab}(j) a_s^3 + O(a_s^4)
\end{eqnarray}
related to Mellin moments of splitting kernels $W_{b\rightarrow a}(x)$ entering 
DGLAP evolution equations \cite{DGLAP} for 
parton distribution densities $f_{a}(x,Q^{2})$  (hereafter $a=\lambda,\,g,\,\phi$ stands for
the spinor, vector and scalar particles respectively):
\begin{eqnarray}
\frac{d}{d\ln {Q^{2}}}f_{a}(x,Q^{2}) &=& \int_{x}^{1}\frac{dy}{y}
\sum_{b}W_{b\rightarrow a}(x/y)\,f_{b}(y,Q^{2}) \, ,
\nonumber \\
\frac{d}{d\ln {Q^{2}}}\tilde{f}_{a}(x,Q^{2}) &=& \int_{x}^{1}\frac{dy}{y}
\sum_{b}\tilde{W}_{b\rightarrow a}(x/y)\,\tilde{f}_{b}(y,Q^{2}) \,.
\label{DGLAP}
\end{eqnarray}
In QCD the anomalous dimensions and splitting kernels are completely known 
up to NNLO
%the next-to-next-to-leading order (NNLO) 
of the perturbation theory
(see \cite{Moch:2004pa,Vogt:2004mw} and references therein). The similar results in the case of $\N$ extended SYM may be obtained from the corresponding QCD results with the help of  Casimir substitutions:  $C_{A}=C_{F}=N_{c}$, $T_{f}n_f={\mathcal N}N_{c}/2$. In addition, in the case of $\N = 2$ and $\N = 4$ SYM we should also account for contributions of scalar particles, see \cite{KL,Kotikov:2003fb} for the case of ${\mathcal N}=4$ SYM. It turns out, however, that the latter calculations accounting for scalar contributions could be completely abandoned in the case of  ${\mathcal N}=4$ SYM. 

Indeed, the expressions for eigenvalues of the anomalous dimension matrix in the ${\mathcal N}=4$ SYM  can be derived directly from the QCD anomalous dimensions without tedious calculations by using a number of plausible arguments. The method elaborated in Ref.~\cite{KL} for this purpose is based on special properties of the integral kernel for the BFKL equation \cite{BFKL,next,KL00} in this model and a new relation between the BFKL and DGLAP equations (see \cite{KL}). First, it turns out that the expression for BFKL kernel in $\N = 4$ SYM posses a property of maximal transcedentality\footnote{See discussion later in this section}. Next, using the mentioned relation of BFKL and DGLAP equations in this model \cite{KL} we may conjecture the similar property of maximal transcedentality for anomalous dimensions \cite{KL}. Moreover, it turns out that eigenvalues of anomalous dimensions matrix in the case of $\N = 4$ SYM are given by the maximal trancedentality part of eigenvalues for QCD anomalous dimension matrix. In NLO approximation this method gives the correct results for anomalous dimensions eigenvalues, which were checked by {\it direct calculations} in Ref.~\cite{Kotikov:2003fb}. Using the results for the NNLO corrections to anomalous dimensions in QCD \cite{Moch:2004pa,Vogt:2004mw} and the method of Ref.~\cite{KL} we have also derived the NNLO eigenvalues of the anomalous dimension matrix in ${\mathcal N}=4$ SYM \cite{KLOV}, which were later confirmed by the integrability predictions in \cite{StaudacherSMatrix}.

Starting from four loops, i.e. above existing QCD results,
the corresponding results for the anomalous dimensions in $\N = 4$ SYM
can be obtained (see \cite{Kotikov:2007cy,Bajnok:2008qj,Lukowski:2009ce}) for example from the long-range asymptotic Bethe ansatz equations together with some additional terms, so-called {\it finite size} or  {\it wrapping
corrections}, coming in agreement with Luscher approach.
\footnote{The three- and four-loop results for the universal  anomalous
dimension have been also reproduced (see \cite{Kotikov:2008pv,Beccaria:2009rw}) by solution of so-called asymptotic long-range Baxter equation.}

\subsection{Maximal transcedentality principle }\label{MethodAD}

As we have already mentioned above in $\N = 4$ SYM there is a way to deduce eigenvalues of anomalous dimensions matrix from QCD results without computing contributions of scalar particles.   This possibility is based on the deep
relation between the DGLAP and BFKL dynamics in the ${\mathcal N}=4$ SYM~
\cite{KL00,KL}.

Indeed, the eigenvalues of the BFKL kernel in $\N = 4$ SYM are
analytic functions of the conformal spin $n$  at least in first two orders of perturbation theory (see Eqs. \eqref{ome1a}, \eqref{7} and (\ref{K3})). Moreover, a detailed inspection  of equations \eqref{7} and \eqref{K3} shows that there is no mixing among special functions of different transcendentality levels $i$
\footnote{
Similar arguments were used also in~\cite{Fleischer:1998nb,Kotikov:2007vr} to obtain analytic results for contributions of some complicated massive Feynman
diagrams without direct calculations.  For the relation of diagrams without sub-divergences and transcedentalities (knots) see \cite{Broadhurst:1995km,Broadhurst:1996kc}.
}.  Indeed, in the framework of ${\overline{\mathrm{DR}}}$-scheme~\cite{DRED},
all special functions entering expression for NLO correction contain only sums of the terms $\sim 1/\gamma^{i}~(i=3)$. More precisely, introducing the
transcendentality level $i$ for eigenvalues $\omega(\gamma)$ of BFKL integral kernel at zero conformal spin  in an accordance with the complexity of the terms in the
corresponding sums
\[
\Psi \sim 1/\gamma ,~~~\Psi ^{\prime }\sim \beta ^{\prime }\sim \zeta
(2)\sim 1/\gamma ^{2},~~~\Psi ^{\prime \prime }\sim \beta ^{\prime \prime
}\sim \Phi \sim \zeta (3)\sim 1/\gamma ^{3},
\]
we see that transcedentality levels of LO and NLO BFKL kernel eigenvalues are given by $i=1$ and $i=3$, respectively.

Now, taking into account the relation between the BFKL and DGLAP equations in $\N = 4$ SYM (see~\cite{KL00,KL}) we may conjecture that similar properties should be also valid for the
anomalous dimensions themselves, i.e. universal anomalous dimensions $\gamma
_{uni}^{(0)}(j)$, $\gamma _{uni}^{(1)}(j)$ and $\gamma _{uni}^{(2)}(j)$ are
assumed to have trancedentality levels $i=1$, $i=3$ and
$i=5$, respectively. An exception could be for the terms appearing at a given
order from previous orders of the perturbation theory and having a lesser trancedentality level. Such contributions could be removed by an approximate finite renormalization of the coupling constant. Moreover, these terms do not appear in the ${\overline{\mathrm{DR}}}$-scheme.

It is known, that in LO and NLO approximations
(accounting for SUSY relation for  QCD color factors $C_{F}=C_{A}=N_{c}$) the
most complicated contributions (with $i=1$ and $i=3$, respectively) are the
same for all LO and NLO eigenvalues of QCD anomalous dimensions matrix \cite{Moch:2004pa,Vogt:2004mw} and also for LO and NLO scalar-scalar anomalous
dimensions in $\N = 4$ SYM \cite{Kotikov:2003fb}. This property allows one to find $\N = 4$ SYM universal anomalous dimensions $\gamma _{uni}^{(0)}(j)$ and $\gamma_{uni}^{(1)}(j)$ without knowing all elements of the anomalous dimensions
matrix \cite{KL}, which was verified by the exact calculations in \cite{Kotikov:2003fb}.

Using above arguments it was concluded in \cite{KL}, that at the NNLO level there is only one possible candidate for $\gamma _{uni}^{(2)}(j)$. Namely, it is the most complicated part of QCD anomalous dimensions matrix
(accounting again for SUSY relation for QCD color factors $C_{F}=C_{A}=N_{c}$). Indeed,  after the diagonalization of QCD anomalous dimensions matrix the most complicated parts (having maximum transcedentality level) of its eigenvalues differ from each other by a shift in their arguments (Lorentz spin) and  the differences are constructed from less complicated terms. The non-diagonal matrix elements of the anomalous dimensions matrix also contain only less complicated terms (see, for example exact expressions for LO and NLO QCD anomalous dimensions 
in Refs.~\cite{Moch:2004pa,Vogt:2004mw} and \cite{Kotikov:2003fb} for the case of $\N = 4$ SYM) and as such they cannot contribute to the most complicated parts of eigenvalues of anomalous dimensions matrix. Summarizing, the most complicated part of NNLO eigenvalues of QCD anomalous dimensions should coincide (up to color factors) with $\N = 4$ SYM universal anomalous dimension $\gamma_{uni}^{(2)}(j)$. 
All these arguments in general apply to singlet anomalous dimensions. Nevertheless, as was shown in \cite{KLOV} the universal anomalous dimension in $\N = 4$ SYM could be also obtained from corresponding QCD results for nonsinglet anomalous dimensions available at that moment. 

We would like to note that the transcedentaly was either used or have been observed also in the study of other quantities. In the same way as above, in 
\cite{Bianchi:2013sta} the universal coefficient function of deep inelastic scattering in the framework of $\N = 4$ SYM was obtained from the most complicated part of the corresponding QCD Wilson coefficients \cite{Vermaseren:2005qc}. For more recent application of
the transcendentality principle see for example the discussion  of energy -energy correlations  \cite{Dixon:2019uzg} and collinear anomalous dimension in \cite{Dixon:2017nat} in the case of QCD and $\N = 4$ SYM. The maximal transcedentality property was also observed in the study of scattering amplitudes (see \cite{Caron-Huot:2019vjl} and references therein) and formfactors \cite{Loebbert:2016xkw}.

\subsection{Universal anomalous dimensions in $\N =4$ SYM}

Let us consider universal\footnote{The considered operators belong to the same $\N = 4$ SYM superconformal multiplet and as such share the same universal anomalous dimension.} anomalous dimensions of the following color and  $SU(4)$ singlet Wilson twist-2 operators 
\begin{eqnarray}
\mathcal{O}_{\mu _{1},...,\mu _{j}}^{g} &=&\hat{S}
G_{\rho \mu_{1}}^{a}{\mathcal D}_{\mu _{2}}
{\mathcal D}_{\mu _{3}}...{\mathcal D}_{\mu _{j-1}}G_{\rho \mu _{j}}^a\,,
\label{ggs}\\
%{\tilde{\mathcal{O}}}_{\mu _{1},...,\mu _{j}}^{g} &=&\hat{S}
%G_{\rho \mu_{1}}^a {\mathcal D}_{\mu _{2}}
%{\mathcal D}_{\mu _{3}}...{\mathcal D}_{\mu _{j-1}}{\tilde{G}}_{\rho \mu _{j}}^a\,,
%\label{ggp}\\
\mathcal{O}_{\mu _{1},...,\mu _{j}}^{\lambda } &=&\hat{S}
\bar{\lambda}_{i}^{a}\gamma _{\mu _{1}}
{\mathcal D}_{\mu _{2}}...{\mathcal D}_{\mu _{j}}\lambda ^{a\;i}\,, \label{qqs}\\
%{\tilde{\mathcal{O}}}_{\mu _{1},...,\mu _{j}}^{\lambda } &=&\hat{S}
%\bar{\lambda}_{i}^{a}\gamma _{5}\gamma _{\mu _{1}}{\mathcal D}_{\mu %_{2}}...
%{\mathcal D}_{\mu_{j}}\lambda ^{a\;i}\,, \label{qqp}\\
\mathcal{O}_{\mu _{1},...,\mu _{j}}^{\phi } &=&\hat{S}
\bar{\phi}_{r}^{a}{\mathcal D}_{\mu _{1}}
{\mathcal D}_{\mu _{2}}...{\mathcal D}_{\mu _{j}}\phi _{r}^{a}\,,\label{phphs}
\end{eqnarray}
where ${\mathcal D}_{\mu }$ are covariant derivatives.
The spinors $\lambda _{i}$ and gauge field tensor $G_{\rho \mu }$ describe gluino and gluon field respectively, while $\phi _{r}$ stand for complex scalar fields.
For all operators in Eqs.~(\ref{ggs})-(\ref{phphs}) $\hat{S}$ denotes  symmetrization of the tensors in the Lorentz indices
$\mu_{1},...,\mu _{j}$ and subtraction of their traces.

The elements of the LO anomalous dimension matrix in the ${\mathcal N}=4$
SYM are given by (see \cite{LN4}):
\begin{eqnarray}
\gamma^{(0)}_{gg}(j) &=& 4
\left( \Psi(1)-\Psi(j-1)-\frac{2}{j}+\frac{1}{j+1}
-\frac{1}{j+2} \right),  \nonumber \\
\gamma^{(0)}_{\lambda g}(j) &=& 8 \left(\frac{1}{j}-\frac{2}{j+1}+\frac{2}{j+2}
\right),~~~~~~~~~~\, \gamma^{(0)}_{\varphi g}(j) ~=~ 12 \left( \frac{1}{j+1}
-\frac{1}{j+2} \right),  \nonumber \\
\gamma^{(0)}_{g\lambda}(j) &=& 2 \left(\frac{2}{j-1}-\frac{2}{j}+\frac{1}{j+1}
\right),~~~~~~~~~~\, \gamma^{(0)}_{q\varphi}(j) ~=~ \frac{8}{j} \,,
\nonumber \\
\gamma^{(0)}_{\lambda \lambda}(j) &=& 4 \left( \Psi(1)-\Psi(j)+\frac{1}{j}-
\frac{2}{j+1}%
\right),~~ \gamma^{(0)}_{\varphi \lambda}(j) ~=~ \frac{6}{j+1} \,, \nonumber \\
\gamma^{(0)}_{\varphi \varphi}(j) &=& 4 \left( \Psi(1)-\Psi(j+1)\right),
~~~~~~~~~~~~~~~\, \gamma^{(0)}_{g\varphi}(j) ~=~ 4
\left(\frac{1}{j-1}-\frac{%
1}{j} \right),  \label{3.2}
\end{eqnarray}
This anomalous dimension matrix could be easily diagonalized and we get simple expression \cite{LN4,KL}:
\begin{eqnarray}
%{\Biggl[D\Gamma D^{-1}\Biggr]}^{N=4}=
D\Gamma D^{-1}=
%_{\mathbf{unpol}} =
\begin{array}{|ccc|}
-4S_1(j-2) & 0 & 0 \\
0  & -4S_1(j) & 0  \\
0 & 0 & -4S_1(j+2)
\end{array}
\, , \nonumber
\end{eqnarray}
%\begin{eqnarray}
%{\Biggl[D\Gamma D^{-1}\Biggr]}^{N=4}_{\mathbf{pol}} =
%\begin{array}{|cc|}
%-4S_1(j-1) & 0  \\
%0  & -4S_1(j+1)
%\end{array}
%\,,\nonumber
%\end{eqnarray}
where 
\begin{eqnarray}
S_1(j) \equiv
\Psi(j+1)-\Psi(1) \equiv \sum_{r=1}^{j}
\frac{1}{r}. \nonumber
\end{eqnarray}
From this we conclude that at LO the universal anomalous dimension for the considered operator supermultiplet is given by
\begin{eqnarray}
\gamma^{(0)}_{uni}(j)~=~-4S_1(j-2) 
%\equiv
%-4\Bigl(\Psi(j-1)-\Psi(1) \Bigr) \equiv -4 \sum_{r=1}^{j-2}
%\frac{1}{r}. 
%$$
\nonumber
\end{eqnarray}
The universal anomalous dimensions at NLO and NNLO could be obtained using the method described in subsection \ref{MethodAD}.  This way up to three loops we get
\footnote{
Note, that in an accordance with Ref.~\cite{next}
 our normalization of $\gamma (j)$ contains
the extra factor $-1/2$ in comparison with
the standard normalization (see~\cite{KL})
and differs by sign in comparison with one from Ref.~\cite{Moch:2004pa,Vogt:2004mw}.}
\cite{KL,KLOV}:
\begin{eqnarray}
\gamma(j)\equiv\gamma_{uni}(j) ~=~ \hat a \gamma^{(0)}_{uni}(j)+\hat a^2
\gamma^{(1)}_{uni}(j) +\hat a^3 \gamma^{(2)}_{uni}(j) + ... , \qquad \hat a=\frac{\alpha N_c}{4\pi}\,,  \label{uni1}
\end{eqnarray}
where
\begin{eqnarray}
\frac{1}{4} \, \gamma^{(0)}_{uni}(j+2) &=& - S_1,  \label{uni1.1} \\
\frac{1}{8} \, \gamma^{(1)}_{uni}(j+2) &=& \Bigl(S_{3} + \overline S_{-3} \Bigr) -
2\,\overline S_{-2,1} + 2\,S_1\Bigl(S_{2} + \overline S_{-2} \Bigr),  \label{uni1.2} \\
\frac{1}{32} \, \gamma^{(2)}_{uni}(j+2) &=& 2\,\overline S_{-3}\,S_2 -S_5 -
2\,\overline S_{-2}\,S_3 - 3\,\overline S_{-5}  +24\,\overline S_{-2,1,1,1}\nonumber\\
&&\hspace{-1.5cm}+ 6\biggl(\overline S_{-4,1} + \overline S_{-3,2} + \overline S_{-2,3}\biggr)
- 12\biggl(\overline S_{-3,1,1} + \overline S_{-2,1,2} + \overline S_{-2,2,1}\biggr)\nonumber \\
&& \hspace{-1.5cm}  -
\biggl(S_2 + 2\,S_1^2\biggr) \biggl( 3 \,\overline S_{-3} + S_3 - 2\, \overline S_{-2,1}\biggr)
- S_1\biggl(8\,\overline S_{-4} + \overline S_{-2}^2\nonumber \\
&& \hspace{-1.5cm}  + 4\,S_2\,\overline S_{-2} +
2\,S_2^2 + 3\,S_4 - 12\, \overline S_{-3,1} - 10\, \overline S_{-2,2} + 16\, \overline S_{-2,1,1}\biggr)
\label{uni1.5}
\end{eqnarray}
and $S_{a} \equiv S_{a}(j),\ S_{a,b} \equiv S_{a,b}(j),\ S_{a,b,c} \equiv
S_{a,b,c}(j)$ are harmonic sums 
\begin{eqnarray}
S_{\pm a}(j) = \sum_{r=1}^{j} \frac{(\pm 1)^r}{r^a},~~
S_{\pm a, \pm b, ...}(j) = \sum_{r=1}^{j} \frac{(\pm 1)^r}{r^a}
S_{\pm b, ...}(r)
\label{hSum}
\end{eqnarray}
%(see Eq. (\ref{FI2}) 
and
\begin{eqnarray}
\overline S_{-a,b,c,\cdots}(j) ~=~ (-1)^j \, S_{-a,b,c,...}(j)
+ S_{-a,b,c,\cdots}(\infty) \, \Bigl( 1-(-1)^j \Bigr).  \label{ha3}
\end{eqnarray}

The expression \eqref{ha3} is defined for positive integer values of arguments
(see~\cite{Kazakov:1987jk,KL,Kotikov:2005gr})
but can be easily analytically continued to real and complex $j$
by the method of Refs.~\cite{Kazakov:1987jk,Kotikov:2005gr,Kotikov:1994mi}. It is interesting to study various limits of the obtained expressions for universal anomalous dimension. 

\subsubsection{The limit $j\rightarrow 1$ }

The limit $j\rightarrow 1$ is important for the investigation of the small-$x
$ behavior of parton distributions (see review \cite{Andersson:2002cf} and references therein). Especially it became popular recently due to
new experimental data for high energy processes studied at LHC. In addition, there are very precise combined experimental data
at small $x$ produced by H1 and ZEUS collaborations at HERA~\cite{Aaron:2009aa}.

Using asymptotic expressions for harmonic sums at $j=1+\omega \rightarrow 1$
(see \cite{KL,KLOV}) the universal anomalous dimension $\gamma_{uni}(j)$ in Eq.~(\ref{uni1}) takes the form 
\begin{eqnarray}
\gamma _{uni}^{(0)}(1+\omega ) &=&\frac{4}{\omega }+{\mathcal O}\Bigl(\omega^{1}\Bigr),
\label{uni4.1} \\
\gamma _{uni}^{(1)}(1+\omega ) &=&-32\,\zeta _{3}+{\mathcal O}\Bigl(\omega^{1}\Bigr),
\label{uni4.2} \\
\gamma _{uni}^{(2)}(1+\omega ) &=&32\zeta _{3}\,\frac{1}{\omega ^{2}}
-232\,\zeta _{4}\,\frac{1}{\omega }
-1120\zeta _{5}+256\zeta _{3}\zeta _{2}
+{\mathcal O}\Bigl(\omega ^{1}\Bigr)  \label{uni4.3}
\end{eqnarray}
in the agreement with the predictions coming from BFKL
equation at NLO accuracy \cite{KL00}.

\subsubsection{The limit $j\rightarrow 4$ }

This limit corresponds to Konishi operator \cite{Konishi:1983hf}, which was used extensively in integrability \cite{integrability-review} and AdS-CFT correspondence \cite{AdS-CFT,AdS-CFT1,Witten} studies recently. The anomalous dimension for Konishi supermultiplet coincides with our expression \eqref{uni1} for $j=4$
\begin{equation}\label{ADKonTL}
\gamma_{uni}(j)\big|_{j=4}
=-6\, \hat a_s + 24\, \hat a_s^2 - 168\,\hat a_s^3\, ,
\end{equation}
which was also confirmed by direct calculation at two \cite{Arutyunov:2001mh,Eden:2005ve,Eden:2005bt,Kotikov:2003fb} and three-loop \cite{Eden:2004ua} orders. Now. there are results for
four, five, six, seven and up to eleven loop corrections to its anomalous dimension
\cite{Bajnok:2008bm,Fiamberti:2007rj,Velizhanin:2008jd},  \cite{Bajnok:2009vm,Arutyunov:2010gb,Balog:2010xa,Eden:2012fe}, 
\cite{Leurent:2012ab,Marboe:2014sya}, \cite{Bajnok:2012bz,Marboe:2016igj} and \cite{VolinPerturbativeSolution,N4SYMQSC4}. Moreover, there are 
 analytical results at strong coupling \cite{Gromov:2011bz,Gromov:2011de,QSCwork} and numerical calculations \cite{Gromov:2009zb,Frolov:2010wt,QSCnumericsN4SYM1,QSCnumericsN4SYM2}
of these anomalous dimensions at any finite coupling.

\subsubsection{The limit $j\rightarrow  \infty  $ }

This limit is related to the study of the asymptotics of structure functions and cross-sections at $x\rightarrow 1$ corresponding to the quasi-elastic kinematics of the deep-inelastic $e p$ scattering. In the limit $j\to \infty $ the expressions for anomalous dimensions \eqref{uni1.1}-\eqref{uni1.5} simplify significantly and takes the form
\begin{eqnarray}
\gamma _{uni}^{(0)}(j) &=&-4\Bigl(\ln j+\gamma _{E}\Bigr)+{\mathcal O}\Bigl(j^{-1}
\Bigr),  \label{uni3.1} \\
\gamma _{uni}^{(1)}(j) &=&8\zeta _{2}\,\Bigl(\ln j+\gamma _{E}\Bigr)+12\zeta
_{3}+{\mathcal O}\Bigl(j^{-1}\Bigr),  \label{uni3.2} \\
\gamma _{uni}^{(2)}(j) &=&-88\zeta _{4}\,\Bigl(\ln j+\gamma _{E}
\Bigr)-16\zeta _{2}\zeta _{3}-80\zeta _{5}+{\mathcal O}\Bigl(j^{-1}\Bigr),
\label{uni3.3}
\end{eqnarray}
where $\gamma _{E}$ is Euler constant.

The strong coupling expansion in this limit could be again studied with the use of AdS/CFT correspondence \cite{AdS-CFT,AdS-CFT1,Witten}. The latter provides us with very interesting strong coupling prediction \cite{Gubser:2002tv} (see also \cite{Kruczenski:2002fb,Makeenko:2002qe,Axenides:2002zf}) for
large-$j$ behavior of anomalous dimensions of twist-2 operators:
\begin{equation}
\gamma (j)=a(z)\,\ln j \,,\qquad\qquad z=\frac{\alpha N_{c}}{\pi } =4\hat a_s
\label{az}\, ,
\end{equation}
where (see Ref.~\cite{Frolov:2002av} for asymptotic
corrections)
\begin{equation}
\lim_{z\rightarrow \infty }a=-z^{1/2}+\frac{3\ln 2}{8 \pi}+{\mathcal O}
\left(z^{-1/2}\right) \,.
\label{1d}
\end{equation}
On the other hand, 
the results for
$\gamma_{uni}(j)$ in Eqs.~(\ref{uni1}) and (\ref{uni3.1})--(\ref{uni3.3})
allow us to find three first terms of the small-$z$
expansion of the coefficient $a(z)$
\begin{equation}
\lim_{z\rightarrow 0}\,a=-z+\frac{\pi ^2}{12}\, z^2-\frac{11}{720}
\pi^4z^3+...\,.
\end{equation}
To resum this series Lev
Lipatov suggested the following equation for the approximate function
$\tilde{a}$~\cite{Kotikov:2003fb}
\begin{equation}
z=-\widetilde{a}+\frac{\pi ^{2}}{12}\,\widetilde{a}^{2}\, ,
\label{approx}
\end{equation}
Solving the latter we may find its  weak-coupling expansion
\begin{equation}
\tilde{a}=-z+\frac{\pi ^{2}}{12}\,z^{2}-\frac{1}{72}\pi ^{4}z^{3}+{\mathcal O}(z^{4})
\end{equation}
and strong-coupling asymptotics
\begin{equation}
\tilde{a} = -\frac{2\sqrt{3}}{\pi} \,z^{1/2} + \frac{6}{\pi^2} + {\mathcal O}
\left(z^{-1/2}\right)
\approx -1.1026\,\,z^{1/2}+0.6079+{\mathcal O}
\left(z^{-1/2}\right).
\end{equation}
It is remarkable, that the predictions for NNLO corrections based on the above simple equation and the  NNLO results for universal anomalous dimension (\ref{uni1.5}) and 
(\ref{uni3.3})) are agree with each other with the accuracy $\sim 10\%$. It means, that this
extrapolation seems to be good for all values of $z$\footnote{Some improvement of (\ref{approx}) can be found in \cite{Bern:2006ew}.}.

The limit $j\to\infty$ could be also studied with the help of Beisert-Eden-Staudacher (BES) integral equation \cite{Beisert:2006ez} for some function $f(x,z)$ related to
$a(z)$ in (\ref{az}) at $x=0$, i.e. $f(0,z)=a(z)$. At weak coupling $z$ this equation gives a possibility to easily calculate as many coefficients 
$c_m$ in the expansion
$$
f(0,z)~= \sum_{m=0} c_m \, \, z^m \, .
$$
as needed. The coefficients $c_m$ respect the principle of maximal transcedentality, 
i.e. $c_m \sim \zeta(2m)$
for $m>0$ (or products of $\zeta$-functions with the sum of indices equal to
$2m$). Moreover, up
to 4-loop order these coefficients are in agreement with the ones, obtained  directly from the calculation of Feynman diagrams \cite{Bern:2006ew,Bern:2007ct}.

The analysis of strong coupling limit $z \to \infty $  of BES equation is less straightforward. Still,  Gupser-Klebanov-Polyakov strong coupling asymptotics $\sim z^{1/2}$ (see the r.h.s. of
(\ref{1d})) was first reproduced from BES equation numerically in  \cite{Benna:2006nd} and then analytically in \cite{Kotikov:2006ts}.  Later the study of strong coupling limit of the function $f(0,z)$ continued in Refs. \cite{Alday:2007qf,Kostov:2007kx,Beccaria:2007tk,Casteill:2007ct,Basso:2007wd,Kostov:2008ax}. In particular, the  authors of \cite{Basso:2007wd} found a way to evaluate subleading
coefficients $\tilde{c}_m$  of the expansion
$$
f(0,z)~= \sum_{m=0} \tilde{c}_m \, \, z^{(1-m)/2}
$$
The first three coefficients
are in agreement with string side calculations performed
in \cite{Gubser:2002tv}, \cite{Frolov:2002av} and \cite{Roiban:2007jf}, respectively.
Moreover, the results of \cite{Basso:2007wd}  are also in agreement with maximum 
transcendentality principe:
%}: 
$\tilde{c}_1 \sim \ln 2$ and
$\tilde{c}_m \sim \zeta(m)$
for $m>1$ (or products of $\zeta$-function with the sum of indexes equal to
$m$).
%, for all coefficients.

\subsection{Anomalous dimensions and integrability}

As we already mention in the section devoted to integrability of BFKL equation in $\N = 4$ SYM the DGLAP equation in this theory is also integrable. Moreover, the integrability property of DGLAP equation extends in this theory in planar limit to all loops. First hints that this might be the case came from the calculation of anomalous dimensions of twist 2 operators in $\N = 4$ SYM \cite{N=4,LN4}. Nevertheless, the real breakthrough in the study of integrability of $\N = 4$ SYM theory started only later with the  work of Minahan and Zarembo \cite{MinahanZarembo}. First, Beisert computed the complete one loop dilatation operator in this theory \cite{Beisert:2003jj} and almost immediately Beisert and Staudacher \cite{N4SuperSpinChain} showed that one-loop dilatation operator in $\N = 4$ SYM may be identified with the hamiltonian of integrable $\mathfrak{psu} (2,2|4)$ super spin chain. This observation together with available techniques for integrable spin chains based on simple Lie algebras \cite{Reshetikhin:1986vd,Ogievetsky:1986hu}  and superalgebras \cite{Saleur:1999cx} allowed authors of \cite{N4SuperSpinChain} to write down one-loop Bethe ansatz equations for the discovered $\N = 4$ SYM spin chain. Next, after a lot of preliminary work the authors of \cite{LongRangeBetheAnsatz} came with  the conjecture for all-loop asymptotic Bethe ansatz equations up to so called dressing factor \cite{BetheAnsatzQuantumStrings}, which was fixed later in \cite{TranscedentalityCrossing} based on crossing equation proposed in \cite{JanikWorldsheetSmatrix}. In what follows we will briefly describe these and subsequents developments in the study of integrability structure of $\N = 4$ SYM.

\subsubsection{Asymptotic Bethe-ansatz and L\"uscher corrections }

The problem of calculation of anomalous dimensions could be reduced to problem of diagonalization of dilatation operator, which is one of the generators of superconformal symmetry of $\N = 4$ SYM. The latter problems in a language of corresponding $\mathfrak{psu} (2,2|4)$ super spin chain is equivalent to finding a spectrum of spin chain hamiltonian upon identification of single trace operators with its states. At one-loop order the hamiltonian of $\mathfrak{psu} (2,2|4)$ is the one of spin chain with only nearest neighbor interactions. Beyond one loop spin chain becomes long range with its all-loop asymptotic Bethe ansatz equations given in \cite{LongRangeBetheAnsatz}. The obtained in \cite{LongRangeBetheAnsatz} Bethe ansatz is asymptotic in a sense that it is valid only when the length of spin chain exceeds interaction order. Another novel feature of $\N = 4$ spin chain is fluctuations of its length, so that the latter is dynamic.  The corresponding $\N = 4$  spin chain S-matrix \cite{StaudacherSMatrix,BeisertDynamicSmatrix} could be decomposed into a product of two $\mathfrak{su}(2|2)$ factors\footnote{The choice of a reference state (vacuum) in a spin chain breaks $\mathfrak{psu} (2,2|4)$ symmetry down to $\mathfrak{su}(2|2)\otimes\mathfrak{su}(2|2)$.}
\begin{equation}
S_{\mathfrak{psu} (2,2|4)} (u_1,u_2) = S_0 (u_1, u_2)\cdot \left(
S_{\mathfrak{su}(2|2)} (u_1,u_2)\otimes S_{\mathfrak{su}(2|2)} (u_1,u_2)
\right)\, , \label{Smatrix-decomposition}
\end{equation}
where $u_1$ and $u_2$ are rapidities of scattering magnons and 
\begin{equation}
S_0(u,v) = \frac{x^{-}(u)-x^+(v)}{x^+(u)-x^-(v)} \frac{1-\frac{g^2}{x^+(u)x^-(v)}}{1-\frac{g^2}{x^-(u)x^+(v)}}\exp (2 i\theta (u,v))\, .
\end{equation}
Here $\theta (u,v)$ is the BES dressing phase \cite{TranscedentalityCrossing} and $\mathfrak{su}(2|2)$ S-matrices could be fixed completely from the symmetry requirements \cite{BeisertDynamicSmatrix}.  The variables $x^{\pm}(u)$ are related to magnon rapidities $u$ through Zhukovsky map
%with
%
\begin{equation}\label{definition x}
x^{\pm}(u)=x(u^\pm)\, ,
\qquad
u^{\pm}=u\pm\frac{i}{2}\, ,
\qquad
x(u)=\frac{u}{2}\left(1+\sqrt{1-4\,\frac{g^2}{u^2}}\right),
\end{equation}

In the case of $\mathfrak{sl} (2)$ sector corresponding to twist 2 operators relevant to DGLAP evolution the long-range asymptotic Bethe ansatz equations for $M$-magnon state take the form
\begin{eqnarray}
\label{sl2eq}
\left(\frac{x^+_k}{x^-_k}\right)^2 ~=~
\prod_{m=1,m\neq k}^M \,
\frac{x_k^--x_m^+}{x_k^+-x_m^-}\,
\frac{(1-g^2/x_k^+x_m^-)}{(1-g^2/x_k^-x_m^+)}\,
\exp\left(2\,i\,\theta(u_k,u_j)\right),
%\nonumber \\
\qquad
 \prod_{k=1}^{\hat{M}} \frac{x^+_k}{x^-_k}=1\, .
\end{eqnarray}
The spin length in this case is equal to two and the number of magnons corresponds to the Lorentz spin of twist 2 operators \eqref{ggs}-\eqref{phphs}.

Once we know the solution for $M$ Bethe roots $u_k$ (magnon rapidities) from the equation above, the asymptotic expression for anomalous dimension (energy of spin chain state) could be found from the formula
\begin{equation}
\label{dim}
\gamma^{ABA}(g)=2\, g^2\, \sum^{M}_{k=1}
\left(\frac{i}{x^{+}_k}-\frac{i}{x^{-}_k}\right) .
\end{equation}
The Bethe ansatz equations in this case can be solved recursively order by order
in coupling constant $g$ at arbitrary integer values of $M$ once the one-loop solution for a given state is known. The later is known from the exact solution of Baxter equation \cite{Eden:2006rx} and is given by Hahn polynomial. The equations for quantum corrections to the one-loop Bethe roots are linear and thus numerically solvable with high precision. To find the expression for anomalous dimensions valid for arbitrary spin values $M$ we may use the principle of
maximal transcendentality \cite{KL} and the fact the function basis for anomalous dimensions in $\N = 4$ SYM is given by harmonic sums. The reconstruction of universal anomalous dimension is then reduced to fixing coefficients in a known basis at a set of fixed integer spin values.  Following this procedure for the four-loop universal anomalous dimension we get \cite{Kotikov:2007cy}
($M=j+2$)
\begin{equation}
\frac{1}{256} \, \gamma^{ABA}_{uni}(j+2) ~=~
\nonumber \\
%& &
4\, S_{-7}+6\, S_{7} + ... +
-\zeta(3) S_1(S_3-S_{-3}+2\,S_{-2,1}),
%\nonumber
\label{fourloop1}
%\end{eqnarray}
\end{equation}
where the symbol $...$ stands for a large set of not shown nested harmonic sums of degree seven. The obtained expression for universal anomalous dimension could be analytically continued also for non-integer spin values\footnote{An explanation for how this is done may be found in \cite{Kotikov:2005gr}.}, in particular to points $M=-r+\omega$ with $r$ - positive integer.  Harmonic sums of degree seven may lead to poles no higher than seventh order in $\omega$.
In fact, it is known that none of the sums in r.h.s. of \eqref{fourloop1}
can produce such a high-order pole except for the two sums
$S_{7}$ and $S_{-7}$. Their residues at $1/\omega^7$ are
of opposite sign. Thus, one immediately sees that the sum of the
two residues does {\it not} cancel. On the other hand, from BFKL studies \cite{KL00,KL} we know that in the vicinity of the Pomeron pole at
$M=-1+\omega$ the four loop anomalous dimensions behaves as
\begin{equation}
\gamma_{uni}(1+\omega) ~\sim ~  1/\omega^4 \, .
\label{fourloop4}
%\end{eqnarray}
\end{equation}
This latter fact tells us that the asymptotic solution for four-loop anomalous dimension is not compete and one should account for so called {\it wrapping corrections}.  

The first account for wrapping or finite size corrections in the case of $\N = 4$ SYM was done with the help of L\"usher type formula \cite{Luscher:1985dn}. The latter consist from two types of contributions: $F$ and $\mu$ terms.  The $F$-term corresponds to interaction of spin chain particles with virtual particle circulating around cylinder, while the $\mu$-term corresponds to the splitting of the spin chain particle in two other particles, which then recombine after circulating around cylinder. To conjecture  L\"uscher type formula in the case of $\N = 4$ SYM authors of  \cite{Bajnok:2008bm,Bajnok:2008qj} performed large 
volume analysis of Thermodynamic Bethe Ansatz  (TBA) equations for the sinh-Gordon model. The conjectured formula for $F$-term contains two contributions: the first one accounts for corrections to ABA and as a consequence to Bethe roots, while the second one is due to propagation of virtual particles around cylinder, which directly changes energy of spin chain state with all $M$ particles of type $a$ as \cite{Bajnok:2008bm,Bajnok:2008qj}:
\begin{equation}
\Delta E(L) = -\sum_{Q=1}^{\infty}\int_{-\infty}^{\infty} \frac{dq}{2\pi} \text{STr}_{a_1} \left[
S_{a_1 a}^{a_2 a}(q, p_1) S_{a_2 a}^{a_3 a}(q, p_2)\ldots S_{a_M a}^{a_1 a}
\right]e^{\tilde{\epsilon}_{a_1} (q) L}\, ,
\end{equation}
where $L$ is spin chain length and the sum is over all possible virtual particles in the theory $Q$, their polarizations $a_1$ and over all possible intermediate states $a_2, \ldots, a_M$. The matrix $S_{b a}^{c a}(q,p)$ describes the scattering of virtual particle of type $b$ and momentum $q$ with the real particle of type $a$ and momentum $p$, while the exponential factor should be understood as propagator of virtual particle. The $\mu$-term contribution could be obtained by taking residues at the poles of $S$-matrices, however in $N=4$ SYM at weak coupling these contributions are absent. It should be noted, that the same L\"usher type corrections were later much more easily obtained by analyzing large volume limit of TBA directly for $\N = 4$ SYM, see for example \cite{TBAN4}.

Accounting for leading wrapping or L\"uscher corrections  \cite{Bajnok:2008bm,Bajnok:2008qj} the full expression for four-loop universal anomalous dimension takes the form
\begin{eqnarray}
\gamma_{uni}(j+2) &=& \gamma^{ABA}_{uni}(j+2) + \gamma^{wr}_{uni}(j+2),
%\label{wr1}\\
\nonumber \\
\frac{1}{256} \, \gamma^{wr}_{uni}(j+2) &=&
\frac{1}{2} \, S^2_{1} \, \biggl[ 2\, S_{-5}+2\, S_{5}
%\nonumber \\
%& &
+ 4\,\left( \HS_{4,1} - \HS_{3,-2} +
      \HS_{-2,-3} -2\, \HS_{-2,-2,1}\right)
\nonumber \\
& &
- 4\, S_{-2}\zeta(3) -5\, \zeta(5)
\biggr]\, ,
\label{wr2}
\nonumber
\end{eqnarray}
which is in full agreement with BFKL predictions (\ref{fourloop4}). Later,  using similar technique and a property of Gribov-Lipatov reciprocity 
(see \cite{Dokshitzer:2006nm,Basso:2006nk,Beccaria:2007pb,Beccaria:2009vt,Beccaria:2009eq,Beccaria:2010tb} and references therein), the five-loop
corrections fo universal anomalous dimensions have been found in 
\cite{Lukowski:2009ce}.

\subsubsection{Thermodynamic Bethe Ansatz and Quantum Spectral Curve}

The systematic method to account for wrapping or finite size corrections is offered by Thermodynamic Bethe Ansatz (TBA). First, the idea that TBA approach could be used for $\N =4$ SYM appeared in \cite{Ambjorn:2005wa}. Within TBA approach the calculation of spin chain spectrum with account for finite size corrections reduces to calculation of partition function of the mirror model. The simplification comes from the fact, that mirror model is considered at large volume corresponding to large imaginary times in the original model. As a consequence, its spectrum is well under control and may calculated using mirror Asymptotic Bethe Ansatz (ABA). Finally, the partition function in the mirror model is calculated using saddle point approximation resulting in integral equations for Bethe roots and hole densities corresponding to the mentioned saddle point. The mirror ABA equations for primary excitations may be obtained via analytical continuation of the ABA equations in the original theory. In addition one should determine all bound states (string complexes) of the mirror theory and calculate their dispersion relations together with scattering matrices using bootstrap method, which should allow to write down generic ABA in mirror model for all excitations including bound states. First, mirror ABA for $\N = 4$ spin chain was described in \cite{Arutyunov:2007tc}. Using the fact that symmetry structure of the scattering matrix  $\mathfrak{su}(2|2)$ \eqref{Smatrix-decomposition} is the same  as that of Hubbard model one can use the TBA solution of the latter \cite{Hubbard1,Hubbard2}. This resulted in string hypotheses formulated in \cite{Arutyunov:2009zu}. Finally, the TBA equations for $\N = 4$ SYM were written in \cite{TBAN4,TBAN4proposal,TBAexcitedstates,TBAMirrorModel}. The further study of TBA equations continued in the simplified form of $Y$ and $T$-systems \cite{YsystemAdS5,TBAfromYsystem,WronskianSolution,SolvingYsystem}

Eventually, the investigation of $\N = 4$ SYM TBA equations has led to their simplified alternative formulation in terms of quantum spectral curve (QSC),
a finite set of functional equations, so called $Q$-system, together with their analyticity and Riemann-Hilbert monodromy conditions \cite{N4SYMQSC1,N4SYMQSC2,twistedN4SYMQSC,N4SYMQSC3,N4SYMQSC4}. The latter is well suited for both numerical solutions for any coupling and state \cite{QSCnumericsN4SYM1,QSCnumericsN4SYM2} as well as to
analytical calculations both at weak and strong coupling \cite{VolinPerturbativeSolution,N4SYMQSC4,QSCwork,cuspQSC,potentialQSC,QSC_StructureConstants,Marboe:2014sya,Marboe:2016igj,KonishiFate}.

In the $\mathfrak{sl}(2)$ sector of $\mathfrak{psu} (2,2|4)$ spin chain relevant to DGLAP evolution the QSC equations (so called $\P\mu$-system) take the form \cite{N4SYMQSC1,N4SYMQSC2,N4SYMQSC3,N4SYMQSC4}:
\begin{align}
\mu_{a b} -\tilde{\mu}_{a b} &= \tilde{\P}_a\P_b - \tilde{P}_b\P_a\, , \\
\tilde{\P}_a &= (\mu\chi)_a^{~b} \P_b\, , \\
\tilde{\mu}_{a b} &= \mu_{a b}^{[2]}\, ,
\end{align}
where 
\begin{equation}
\chi^{a b} = \begin{pmatrix}
0 & 0 & 0 & -1 \\ 0 & 0 & +1 & 0 \\ 0 & -1 & 0 & 0 \\
+1 & 0 & 0 & 0
\end{pmatrix}\, ,\quad (\mu\chi)_a^{~b}\equiv \mu_{a c}\chi^{c b}
\end{equation}
and indexes $a, b$ take values $1, 2, 3, 4$. $\mu_{a b}$ is antisymmetric matrix constraint by
\begin{equation}
\mu_{12}\mu_{34} - \mu_{13}\mu_{24} + \mu_{14}\mu_{23} = 1\, , \quad \mu_{14} = \mu_{23} 
\end{equation}
and all functions $\P_a$ and $\mu_{a b}$ are functions of spectral parameter $u$.
The functions $\P_a$ have only one Zhukovsky cut ($u\in [-2g, 2g]$) on the defining (physical) sheet, while functions $\mu_{a b}$ have infinitely many branch points at positions $u = \pm 2 g + i\mathbb{Z}$. Next, $f^{[n]}(u) = f(u+\frac{i n}{2})$ and $\tilde{f}(u)$ denotes analytic continuation of function $f (u)$ around one of its branch points on the real axis located at $-2g$ and $2g$. The boundary conditions for the above Riemann-Hilbert problem are specified by the large $u$ asymptotics of $\P_a$ and $\mu_{a b}$ functions:
\begin{align}
& \P_1 \simeq A_1 u^{-\frac{L+2}{2}}\, , \P_2 \simeq A_2 u^{-\frac{L}{2}}\, , \P_3\simeq A_3 u^{\frac{L-2}{2}}\, , \P_4\simeq A_4 u^{\frac{L}{2}}\, , \nonumber \\
& \mu_{12}\sim u^{\Delta - L}, \,  \mu_{13}\sim u^{\Delta - 1}, \, \mu_{14}\sim u^{\Delta}, \,  \mu_{24}\sim u^{\Delta + L} , \,  \mu_{34}\sim u^{\Delta + L}
\end{align}
with
\begin{align}
A_1 A_4 &= \frac{[(L-S+2)^2 - \Delta^2][(L+S)^2 - \Delta^2]}{16 i L (L+1)}\, , \nonumber \\
A_2 A_3 &= \frac{[(L+S-2)^2-\Delta^2][(L-S)^2 - \Delta^2]}{16 i L (L-1)}\, .
\end{align}
Here $L$ is the twist, $S$ - spin and $\Delta$ is conformal dimension of the $\mathfrak{sl}(2)$ operator under consideration. Choosing appropriate ansatz for $\P_a$ functions its possible to solve this Riemann-Hilbert problem up to in principle arbitrary order of perturbation theory \cite{VolinPerturbativeSolution}. Moreover, using computed values of anomalous dimensions at fixed spin values and the basis of binomial harmonic sums\footnote{The latter is possible with the use of  Gribov-Lipatov reciprocity.} it is possible to reconstruct universal anomalous dimension at the required order of perturbation theory. This way the universal anomalous dimensions at six and seven loop order were obtained in Refs. \cite{Marboe:2014sya} and \cite{Marboe:2016igj} correspondingly. 

Finally, as we already noted the use of QSC is not limited by weak coupling regime and the authors of \cite{QSCwork} where able to obtain analytical results for anomalous dimensions of short $\mathfrak{sl}(2)$  operators at first four orders in strong coupling expansion.

\section{Conclusion}

We tried to briefly review the present knowledge of the properties of BFKL and DGLAP equations in the case of $\N = 4$ SYM. A lot of current results in $\N = 4$ SYM were either obtained with the use of {\it transcendentality principle} or confirm the later property of $\N = 4$ SYM model. Obtained originally as a property of BFKL kernel eigenvalues at NLO, the maximum transcedentality property was later observed also for universal anomalous dimensions up to very high order of perturbation theory, scattering amplitudes, form-factors and correlation functions in $\N = 4$ SYM.  

First, the transcedentality principle was used to extract the results for $\N = 4$ SYM anomalous dimensions from the known QCD results at first three orders of perturbation theory. Next, the results for four and five loop universal anomalous dimensions were obtained  from the asymptotic long-range Bethe equations with account for the so-called {\it wrapping corrections} in L\"usher approach. Still, the transcedentality property was used again to construct a basis of harmonic sums, in terms of which the result is written. To get six and seven loop result the quantum spectral curve approach was used instead of asymptotic Bethe equations and L\"usher corrections. 

AdS/CFT correspondence and all-loop integrability of DGLAP and BFKL equations
gave us a unique opportunity to study Pomeron solution and anomalous dimensions of different types of operators at strong and in general at finite coupling.
The most suitable approach for later studies at a moment is offered by a variant of Thermodynamic Bethe Ansatz known as Quantum Spectral Curve approach.

%\section{Conclusion}

%\bibliographystyle{plain}
%\bibliographystyle{ieeetr}
%\bibliographystyle{hieeetr}
%\bibliographystyle{dinat}
%\bibliographystyle{abbrv}
%\bibliographystyle{hep}
\bibliography{litr}

\bibliographystyle{hieeetr}
%\bibliography{litr}

\end{document}